# High yield synthesis and liquid exfoliation of two-dimensional belt like hafnium disulphide

*Harneet Kaur[1,*], Sandeep Yadav[2,§], Avanish K. Srivastava[1], Nidhi Singh[1], Shyama Rath[3], Jörg J. Schneider[2,§], Om P. Sinha[4] and Ritu Srivastava[1,*]*

[1]National Physical Laboratory, Council of Scientific and Industrial Research, Dr. K. S. Krishnan Road, New Delhi 110012, India.

[2]Technische Universität Darmstadt, Eduard-Zintl-Institut für Anorganische und Physikalische Chemie L2 I 05 117, Alarich-Weiss-Str 12, 64287 Darmstadt, Germany.

[3]Department of Physics and Astrophysics, University of Delhi, Delhi 110007, India.

[4]Amity Institute of Nanotechnology, Amity University UP, Sector 125, Noida, Uttar Pradesh 201313, India.

*address for correspondence: harneet@mail.nplindia.org

ritu@nplindia.org

§Material synthesis and characterization

ABSTRACT

Producing monolayers and few-layers in high yield with environment-stability is still a challenge in hafnium disulphide ($HfS_2$), which is a layered two-dimensional material of group-IV transition metal dichalcogenides, to reveal its unlocked electronic and optoelectronic applications. For the first time, to the best of our knowledge, we demonstrate a simple and cost-effective method to grow layered belt-like nano-crystals of $HfS_2$ with surprisingly large interlayer spacing followed by its chemical exfoliation. Various microscopic and spectroscopic techniques reveal these as-grown crystals exfoliate into single or few layers in some minutes using solvent assisted ultrasonification method in N-Cyclohexyl-2-pyrrolidone. The exfoliated nanosheets of $HfS_2$ exhibit an indirect bandgap of 1.3 eV with high stability against ambient degradation. Further, we demonstrate that these nanosheets holds potential for electronic applications by fabricating field-effect transistors based on few layered HfS$_2$ exhibiting field-effect mobility of 0.95 cm$^2$/V-s with a high current modulation ratio (Ion/Ioff) of 10$^4$ in ambient. The method is scalable and has potential significance for both academy and industry.

**Introduction**

Recently, two-dimensional (2D) layered inorganic materials have generated significant interest among the research community [1-3]. A lot of research activity in the past few years has demonstrated the potential of these materials in the field of high speed electronics and optical detectors [3-4], energy generation and storage [5-6], chemical and gas sensors [7-8] and in bio-sensing [9-10]. Currently, this 2D class consists of many families, the simplest being honeycomb lattices, "graphene" which is an excellent conductor [11] on one end and "hexagonal-boron nitride (h-BN)" which is an excellent insulator [12] on the other extreme. Between these two extremes, there exists a wide family of 2D class, "transition-metal oxides and dichalcogenides (TMDs)" consisting of 88 members [13]. Depending upon the group of TMDs, it consists of semi-metals, semiconductors and superconductors with strong covalent coupling in the monolayers and weak interlayer coupling [13]. The most explored group among them is VIB TMDs consisting of semiconducting layered materials like molybdenum disulphide (MoS$_2$) belonging to D$_{6h}$ point-group symmetry with 2H structure [14-16]. But theoretically, it has been predicted that group IVB TMDs belonging to D$_{3d}$ point-group with 1T-structure have superior opto-electronics than the group VIB ones [17]. Despite this fact, very sparse work has been reported on group IVB TMDs.

A group of researchers recently demonstrated phototransistors based on hafnium disulphide (HfS$_2$), which is a group IVB TMD, exhibiting high photosensitivity [18]. Even top-gated transistors with HfS$_2$ as a conducting channel material were also demonstrated, offering high current modulation ratios [19-20]. However, most of this recent work on HfS$_2$ [18-20] was based on micromechanical cleavage technique from single crystals. This method is reliable only at nano-scale level and not for large-area deposition. Even bottom-up approaches like chemical-vapor deposition are still blank in producing a level growth of $HfS_2$ atomic layers [21]. DFT calculations have predicted a large interlayer interaction energy among the $HfS_2$ layers compared to $MoS_2$ making it difficult to exfoliate this material [22]. But to realize the so far unexplored applications of $HfS_2$ in the 2D regime, it is crucial to grow its high quality crystal which can be exfoliated easily by using simple wet-chemical routes for its high-yield production. In the past, Hafnia and Zirconia as well as $ZrS_3$ nano-belts have been reported to be accessible by high temperature routes [23-24]. Also, wet-chemical approaches have been proved to be very promising in producing high-yield of monolayers or few-layered nanosheets of 2D materials dispersed in solvents [25-28]. These methods apart from being versatile are cost-effective, simple and scalable. One such well established method for obtaining layered materials among those is solvent assisted ultrasonification [25]. Using solvents like N-methyl-2-pyrrolidone, various TMDs like $MoS_2$, $MoSe_2$, and $WS_2$ etc. has been exfoliated [25]. Even a new member of this 2D class "phosphorene" has been exfoliated using the same approach [29]. The inexpensive realization of large-area electronics using these exfoliated suspensions in techniques like Langmuir-Blodgett assembly [30-32] or inkjet printing [33] by depositing these nanosheets on any kind of solid support often provides an advantage over other bottom-up approaches like chemical-vapor deposition. Also, these free-standing nanosheets in solvents can be readily mixed with proteins or organic materials to make functional hybrid structures having applications in biosensing [34], organic light emitting diodes [35] or photovoltaic applications [36].

In this work, we demonstrate a chemical-vapor transport route to produce agglomerates of layered belt like shiny crystals of $HfS_2$ [37] and its exfoliation by chemical route for the first time, to the best of our knowledge. The as-prepared belt-like crystals possess a slightly extended interlayer spacing, offering the possibility of exfoliation. Indeed by using solvent assisted ultrasonication in N-Cyclohexyl-2-pyrrolidone (CHP), under ambient conditions, yields stable dispersions of exfoliated $HfS_2$ as single layer up to few layers. A high yield of monolayers and a few layers in solvent CHP has been achieved under ambient conditions. The stable dispersions of exfoliated $HfS_2$ nanosheets shows the presence of an indirect bandgap of 1.3 eV and are found to be stable against surface degradation for many days unlike mechanical exfoliated flakes [22]. Finally, we demonstrate a bottom gate field-effect transistor (FET) with exfoliated $HfS_2$ as a conducting channel material, exhibiting a field-effect mobility of 0.95 $cm^2$/V-s with $SiO_2$ as a dielectric medium. The impressive gate tunability of the drain current as well as a high current modulation $I_{on}/I_{off}$ ratio greater than $10^4$ allows this material to be used for opto-electronic applications.

**Methods**

Crystal Growth:

Single crystals of $HfS_2$ were prepared by the chemical vapor transport (CVT) method. In this work, stoichiometric proportions of hafnium foil (99.9%, 0.25 mm thick, chemPUR GmbH) and sulphur powder (99.998%, Sigma Aldrich) with 5 mg iodine (99.8%, Sigma Aldrich) as a transport agent, added in an evacuated (approx. 3 x $10^{-3}$) clean quartz ampule having inner diameter 1 cm and 0.25 mm wall thickness. The ampule was loaded in the CVD furnace (HORST GmbH) and temperature was increased to 925°C at the constant rate of 2°C/min and kept at 925 °C for 2 hours, followed by a slow cooling at the rate of 1°C/min to a temperature of 825 °C and continued the growth process for four days. After four days, furnace was cooled down to the room temperature and red metallic/orange crystals were obtained and used for characterization. X-ray diffraction (XRD), Raman spectroscopy, scanning electron microscopy (SEM), energy dispersive analysis of X-Rays (EDAX) and transmission electron microscopy (TEM) was performed using Rigaku miniFlex 600, Jobin-Yuon Horiba ( HR-800Raman, 488 nm argon laser), Philips XL-30 FEG, XL30-939-2.5 CDU-LEAP-detector and Tecnai G2 F30 STWIN, TEM respectively.

Exfoliation:

As-prepared $HfS_2$ crystals (50 mg) were dispersed in CHP (10ml, Sigma-Aldrich) and kept in a sealed vial of volume 20 ml. Further, these sealed vials were wrapped with parafilm before placing it into elmasonic TI-H5 bath sonicator. The dispersion was sonicated in DI water for 2 hours at 25 KHz. Temperature of DI water in ultrasonic bath was maintained at 30°C. The colour of the suspension gradually changes from light transparent yellow to dark orange-yellow in two hours, yielding a suspension of few-layers $HfS_2$ nanosheets in CHP. The suspension was centrifuged at 3000 r.p.m for 45 min in a spinwin micro centrifuge (Tarsons) to remove the un-exfoliated material and the supernatant was decanted, and subjected to UV-Visible absorption spectroscopy using Shimadzu, UV-2401. Further, an additional centrifugation at a high speed of 13,000 r.p.m (120 min) is employed to suspend the exfoliated nanosheets in toluene. The supernatant resulted from this high speed centrifugation consists of transparent CHP and a precipitate containing exfoliated nanosheets of $HfS_2$. The precipitate was extracted and mixed in toluene by mild shaking. The mixed suspension of exfoliated nanosheets in toluene was again subjected to a high speed centrifugation at 13,000 r.p.m. (30 min) to remove any traces of CHP. This process is repeated thrice and final precipitate was suspended in toluene (concentration ~0.01 mg/ml) and used for characterizations by making films on $SiO_2$/Si substrates and TEM grids by drop-cast method.

FET fabrication:

The devices were fabricated on doped silicon substrates ($\sim 3 \times 10^{17}$ cm$^{-3}$) with silicon dioxide (230 nm) as a dielectric medium with patterned gold electrodes of channel length 2 microns (Fraunhofer). To make the devices, large sheets of $HfS_2$ were size separated by centrifugation of the toluene suspension at a low speed of 4,000 r.p.m. for 45 min. This resulted in a precipitate consisting of larger few layer sheets of $HfS_2$ which were diluted with toluene and drop-cast onto the gold patterned chips to make devices. The devices were annealed in vacuum ($\sim 10^{-5}$ mbar) at 120$^o$C for six hours before room temperature measurements were taken using keithley 4200 equipped with semiconductor characterization system (Summit 11000M, probe station). Atomic force microscopy (Pro P47 SOLVER, NT-MDT) in tapping mode was used to measure the parameters and thickness of the as-formed device.

**Results and Discussion**

The HfS2 nano-crystals were grown using chemical vapor transport method. Hafnium metal foil (0.25 mm thick) and sulfur powder were used as precursors and iodine as a transporting agent. The details for the growth were discussed in detail in the method section. Various microscopic and spectroscopic techniques have been employed to confirm the morphology, crystal structure and stoichiometry of the as-prepared crystal. Figure 1a shows the as-grown crystals in the quartz ampule. The crystals was found to be made up of small reddish-orange elongated one dimensional belt shaped structures as found on the walls of quartz ampule (Figure 1b). To confirm the morphology, scanning electron microscopic (SEM) was used, revealing the growth of these elongated crystals on the surface of hafnium metal foil (Figure 1c). Individual crystals have lengths larger than hundred of microns and a width of a few microns (Figure 1d). Further, high magnified SEM images (Figure 1e,f,g) confirms the presence of layered morphology in each of these belt like crystalline structures. The presence of hafnium and sulfur elements (Hf and S) in the crystal was confirmed by energy dispersive analysis of X-rays (EDAX) at various spots as shown in the inset of Figure 1h and X-ray photoelectron spectroscopy (XPS) of a micromechanically cleaved crystal on silicon. The AFM image of the cleaved crystal is shown in Figure S1 in the Electronic Supplementary Material (ESM). The XPS peaks (Figure 1i) are observed at 17.7 eV and 16.1 eV corresponds to the hafnium states 4f5/2 and 4f7/2 respectively whereas peak at 160.7 eV and 162 eV corresponds to the sulfur states 2p1/2 and 2p3/2 respectively [21]. The stoichiometry of *Hf:S* as calculated by the area under the XPS peaks was found to be 1:2.1 which is close to the atomic percentage 1:2 as observed in the EDAX in the inset of Figure 1h.

The crystal structure of HfS2 is confirmed by X-ray diffraction and Raman spectroscopy. As shown in Figure 2a, the diffraction pattern from HfS2 matches well with *JCPDS* card No. 28-0444 confirming its CdI2-type layered structure [14] belonging to the space group P̄3m1. The high intensity peak at 2Θ = 14.920o corresponding to (001) *hkl* plane has a narrow full-width-at-half-maximum (<0.2o) and is slightly shifted from its value 15.132o (inset of Figure 2a). No such shift has been observed for the other *hkl* planes. The high intensity of (001) *hkl* plane indicates the presence of vertically oriented growth [21] with high crystallinity, confirming formation of elongated crystals. The shift in the 2Θ value indicates a change in the *d* spacing between the (001) planes from 5.85 Å to 5.93 Å. This has resulted in a 1.36 % of increase in the interlayer spacing compared to the standard hexagonal plates of HfS2. This may be attributed to the preferred growth conditions resulting in the formation of elongated layered belts instead of hexagonal plates. The Raman spectra of a single belt like HfS2 shown in Figure 2b further confirms the structure. The Raman modes at 255 cm-1, 318 cm-1 and 337 cm-1 assigned to Eg, Eu (LO) and A1g modes respectively [38] are congruent with the 1T phase of HfS2. The Eg and Eu (LO) modes corresponds to the in-plane vibrations, whereas A1g mode corresponds to out of plane vibration [38]. It has been reported that A1g Raman mode is highly intense where as Eg and Eu (LO) modes are weak for standard hexagonal plates of HfS2 [22,38]. However, the enhancement of the in-plane vibrations modes Eg and Eu (LO) with a suppression of the out-of-plane vibrations (A1g) in case of nanobelts suggests quantum confinement effects [39-40] or change in the interlayer spacing between the multilayers of HfS2 in 1T-phase, which is well supported by our XRD results effecting Raman intensity. High resolution transmission electron microscopy (HRTEM) is also employed to calculate the distance between the atomic planes. Figure 2c shows the TEM image of a belt like HfS2. The inset represents the change in contrast with an increase in the number of layers.

The atomic-scale image revealed the presence of hafnium and sulfur atoms arranged in 1T-phase of HfS2 with interlayer spacing of 0.6 nm. This result is consistent with the value of *d* spacing for (001) plane as calculated by XRD. Therefore, the formation of highly crystalline HfS2 with a slight extension in interlayer spacing is confirmed. Further, the distance between the two hafnium atoms was calculated to be 0.3 nm which matches well the unit cell parameter of HfS2 (a = 0.36 nm). A corresponding selected-area-electron diffraction pattern (SAEDP) from a belt like structure shows the presence of a set of important atomic planes 010 and 001 along the [100] zone axis of a tetragonal crystal structure of HfS2 in reciprocal space, as displayed in Figure 2c.

For delaminating single layers of HfS2 from individual crystals, a solvent assistant ultrasonication method was employed. A crystal was suspended in CHP and exposed to ultrasonification (power~500 W). These waves create cavitation bubbles, collapsing into high energy jets, resulting in the breaking of van der Waals forces in layered materials. However, it has been reported [25] that the surface energy of the solvent used, should be similar to the energy required to avoid re-aggregation of its layers. Since, CHP belongs to the group of amide solvents like N-methyl-2-pyrrolidone which has been frequently used to exfoliate other 2D materials in ambient oxygen [25]. Therefore, this solvent was chosen to carry out the exfoliation process. The ultrasonification of the as-prepared material in CHP yields a transparent yellow-orange dispersion as seen in the inset of Figure 3a in a time span of just a few minutes. The un-exfoliated material was removed by centrifugation at 3,000 r.p.m for 45 min to yield a suspension of few-layered HfS2 sheets in CHP (see the method section for complete details). Since CHP is a high boiling point solvent which cannot be evaporated at room temperature, an additional high speed centrifugation at 13,000 r.p.m (120 min) is employed to separate out the exfoliated nanosheets of HfS2 from CHP. This results in the extraction of exfoliated nanosheets by discarding the transparent CHP supernatant. The extracted sheets were suspended in toluene by mild shaking and used for further characterizations. SEM images on the drop-cast films revealed (Figure 3a) the beginning of exfoliation by breaking the elongated crystals into thin layered structures which is further confirmed by TEM (inset Figure 3a) revealing the presence of few layered HfS2 sheets. As the sonication time increases, the transparency of the suspension decreases with the increase in the brightness of the suspension (inset of Figure 3b), which suggests increase in the concentration of a few-layered HfS2 sheets in CHP. SEM (Figure 3b) and TEM images (inset Figure 3b) confirms the presence of ultra-thin layers HfS2 confirming the completion of exfoliation process within 120 minutes. The atomic scale TEM image (Figure 3c) of an ultrathin sheet (inset of Figure 3b) shows the top-view of 1T phase of HfS2. A clear centered hexagonal lattice structure is observed without the need of Fourier transform of the image where each hafnium atom is surrounded by six sulfur atoms. Also, the interplanar spacing between the hafnium planes i.e; 0.4 nm corresponding to (100) *hkl* plane and 0.2 nm corresponding to (101) *hkl* planes is close to the *d* spacing 0.32 nm for (100) *hkl* planes and 0.18 nm for (101) *hkl* planes as calculated by the XRD. Due to the different calibrations, instrument errors and limits of resolution of TEM and XRD, a difference in second decimal place is presumably expected. However, the lattice structure appears to be intact over wide regions suggesting that HfS2 can be exfoliated like other 2D materials without introduction of defects. Figure 3d represents the SAEDP of the HfS2 nanosheet (inset of Figure 3b) revealing the presence of important (*hkl*) planes in reciprocal space. This confirms the high crystallinity and the perseverance of pristine phase of HfS2 during exfoliation. To assess the lateral dimensions of exfoliated HfS2 sheets, we performed statistical FE-SEM analysis, confirming a dominating lateral size distribution associated with L~1μm (Figure 3e). The thickness dependent studies using AFM as a function of centrifuge speed and distribution of thickness of exfoliated flakes is discussed in the section S2 and S3 respectively of ESM. To further characterize the HfS2 exfoliated dispersion, UV Visible absorption spectroscopy is employed. Figure 3f shows the absorbance of HfS2 as a function of sonication time. The increase in absorbance clearly suggests the increase in concentration of HfS2 nanosheets in CHP solvent with the increase in sonication time. Further, the plot of $(\alpha h v)^{1/2}$ versus $hv$ (inset of Figure 3f) where α, h and v represents absorbance, Planck's constant, and frequency respectively, shows linearity confirming its an indirect band-gap material. The indirect bandgap as calculated from the intercept of the plot increased from 0.9 eV to 1.3 eV with the increase in sonication time suggesting that the indirect bandgap increases with the decrease in layer number. Our results are consistent with the DFT simulated bandgap of HfS2 as recently reported by the other group [22].

    The structural integrity of the exfoliated sheets is confirmed by Raman spectroscopy. The Raman spectrum shown in Figure 4a represents the characteristic Eg, Eu(LO) and A1g phonon modes of HfS2 nanosheet (inset of Figure 4a). The Eg and Eu(LO) modes which correspond to the vibrations in the basal plane are found highly intense while the A1g phonon mode which corresponds to out of plane basal vibrations, is of low intensity. This is also observed in case of bulk nanobelts (Figure 2b). The enhancement in the intensity of the in-plane vibrations (Eg and Eu) may be due to the extended interplanar spacing (as observed in TEM, Figure 3c) resulting in the oscillations of atoms more freely while decrease in intensity of out-of plane vibrations (A1g) can be attributed to dominating quantum confinement effects [39-40]. Further, it has been reported that the surface of HfS2 readily gets oxidized in

presence of ambient oxygen increasing its thickness by 250 % due to the penetration of oxygen atoms inside the layers within a time span of 24 hours [22]. This significantly decreases the Raman intensity of the phonon modes [22]. However, we found that exfoliation in CHP provides shielding against the degradation of sheets due to ambient oxygen for several days as compared to the ones exfoliated in NMP and dimethylformamide (see section S4 in ESM). Our results are consistent with the results observed for phosphorene nanosheets exfoliated in CHP [41]. The Raman spectra taken after a week of ambient exposure, as shown in Figure 4a, showed the persistence of highly intense $E_g$ and $E_u$ modes. Also, AFM is a reliable tool to study the degradation of the surface. Figure 4b represents the AFM images of HfS2 nanosheet continuously scanned for 6 days. No change in the thickness or any degradation on the surface is observed. This suggests sheets exfoliated in CHP provide stability against ambient degradation.

Furthermore, we investigated the electronic applications of exfoliated HfS2 nanosheets by fabricating field-effect transistors with silicon-dioxide (230 nm) as a bottom-gate dielectric. Figure 4c shows the AFM image of the fabricated FET. A semiconducting channel is build by an HfS2 nanosheet having a thickness of 18 nm between two gold electrodes (thickness 30 nm each) acting as source and drain respectively, providing a channel length (L) of 2 microns. The channel width (W) is found to be 1.2 microns. The source electrode was grounded and the drain electrode was given a positive bias with respect to source (VDS). The back gate voltage (VBG) was applied to the doped silicon substrate. The output characteristic of the fabricated FET is shown in Figure 4d. A clear saturation of drain current (IDS) was observed corresponding to each gate bias voltage which was tuned from -2V to 10 V. Figure 4e shows the transfer characteristic of the FET. The left and right y-axis represents the linear and logarithmic scale respectively for the variation of IDS with VBG (x-axis) for a fixed VDS of 3V. The device exhibits uni-polar n-type transport behavior. A maximum drain current of 0.24 µA/µm is obtained at a gate voltage of 20 V. The FET device exhibits a high current on-off modulation ratio larger than 10,000 at room temperature when gate voltage is varied from 0V to 20 V. Further, low-field effect mobility was extracted from the relation: $\mu = g_m C_g^{-1} V_{DS}^{-1} L/W$, where $g_m$ represents transconductance (calculated from the slope of IDS versus VBG), $C_g$ represents gate capacitance ($\varepsilon_0 \varepsilon_r/d = 1.501 \times 10^{-4}$ F/m2). The mobility was coming out to be 0.95 cm$^2$/V-s. This low value of mobility with a high on/off current modulation ratio at room temperature agrees well with the reported values of other groups [19-20] on micromechanical exfoliated flakes of HfS2. Hence, solvent assisted ultrasonication method produce nanosheets of HfS2 which retain their pristine semiconducting properties as shown by the high quality mechanical exfoliated sheets. However it was found that for thicker exfoliated flakes, the current on/off modulation ratio falls as discussed in section S5 of ESM, limiting the electronic applications of thicker nanosheets similar to other 2D materials.

**Conclusion**

In summary, we have demonstrated a systematic route for the high yield production of HfS2 layered nano-crystals followed by providing a simple solvent assisted ultrasonication process for their exfoliation in CHP to produce their ultra-thin layers in huge quantities. The ultra-thin nanosheets were found to retain their pristine surface without any ambient degradation for many days unlike mechanical exfoliated flakes. Further, these nanosheets were found to be as competent as other two dimensional materials in optoelectronic applications as demonstrated by fabricating HfS2-FET providing a high current modulation of over 10000 at room temperature. Our results open avenues for the inexpensive production of layered HfS2 in high yield to reveal the unexplored applications of this layered material.

FIGURES

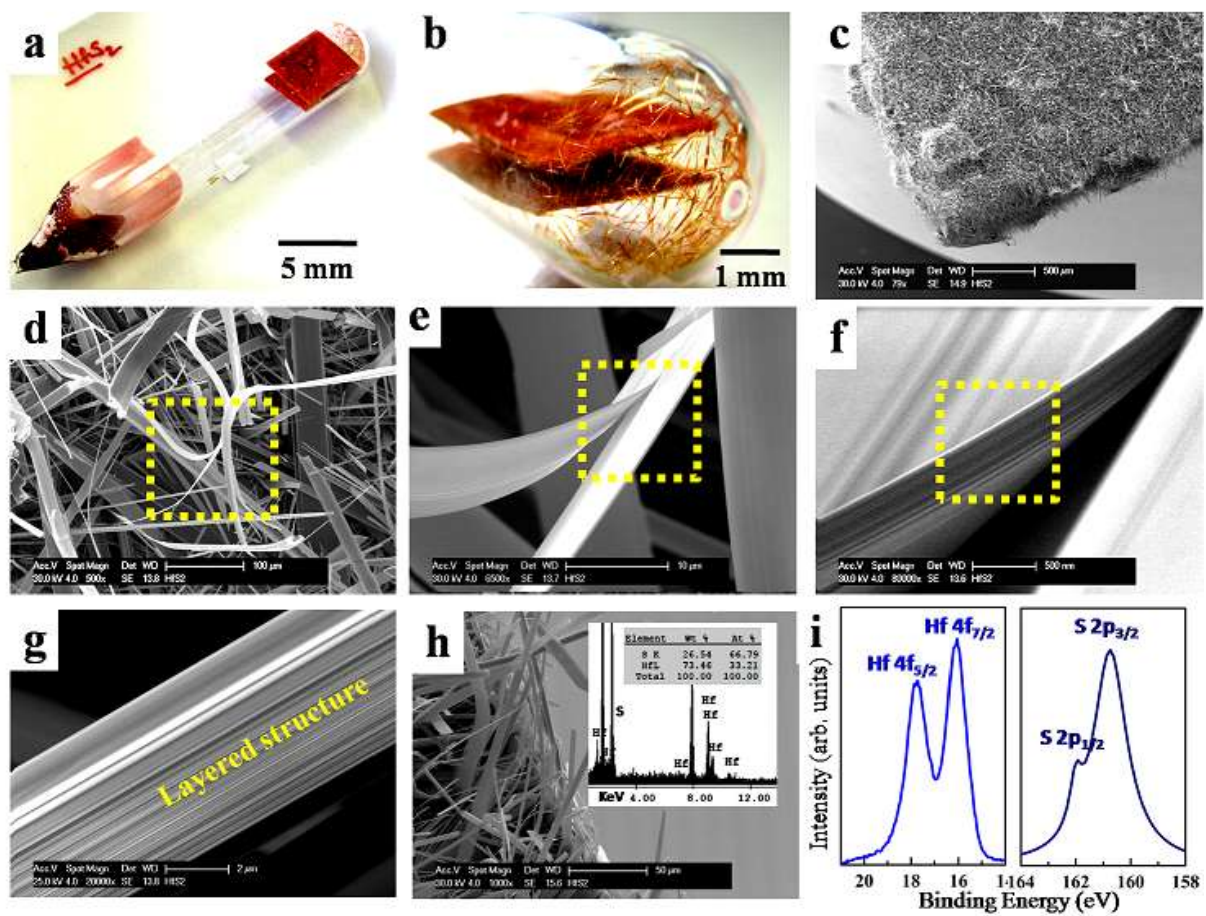

**Figure 1. Morphology and stoichiometry of the as-grown crystals.** (a,b) photographs of the as-grown crystal on the surface of *Hf* foil in the quartz ampule. (c,d) SEM images of the crystal confirming belt like elongated growth. (e,f,g) SEM images confirming the presence of layered structure in each belt (yellow box shows the magnified region in the consecutive image). (h) SEM image of the nanobelts present at edge of the *Hf* foil where EDAX is taken (inset) confirming the atomic percentage of *Hf:S* to be 1:2. (i) XPS spectra of the micromechanically cleaved nanobelts (section S1, ESM) on the surface of SiO2/Si substrate.

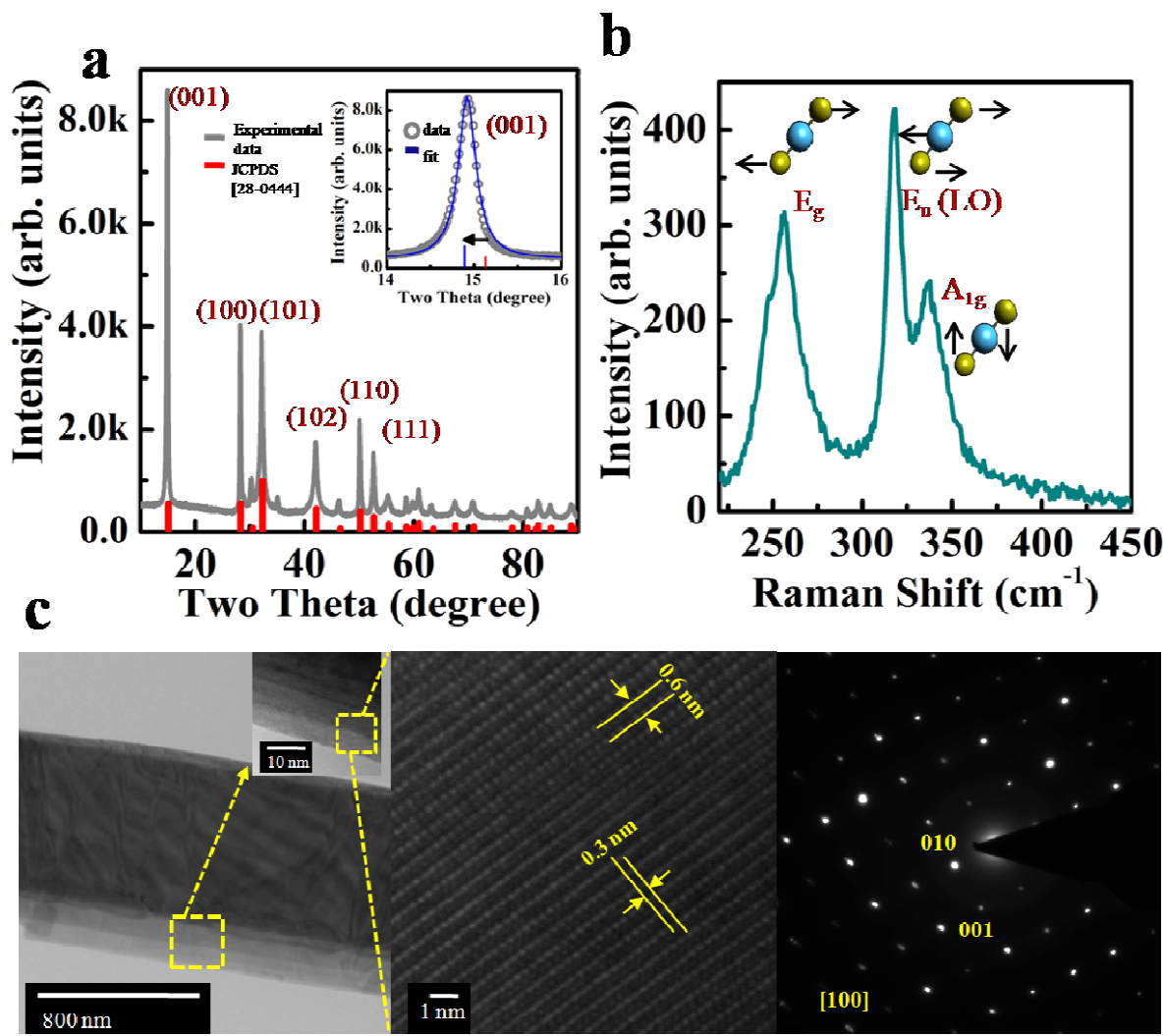

**Figure 2. XRD, Raman and TEM analysis of HfS2 nanobelt.** (a) XRD of the crystal (grey) and JCPDS card 28-0444 (red). Inset shows the shift of XRD peak for (001) *hkl* plane. (grey-experimental data, blue-Lorentzian fit). (b) Raman spectra of a belt like HfS2 crystal. (blue sphere: *Hf* atom, yellow sphere: *S* atom). (c) TEM image of HfS2 belt like crystal, its magnified image (inset), atomic scale images revealing presence of 1T-HfS2 phase and selected area electron diffraction pattern (SAEDP).

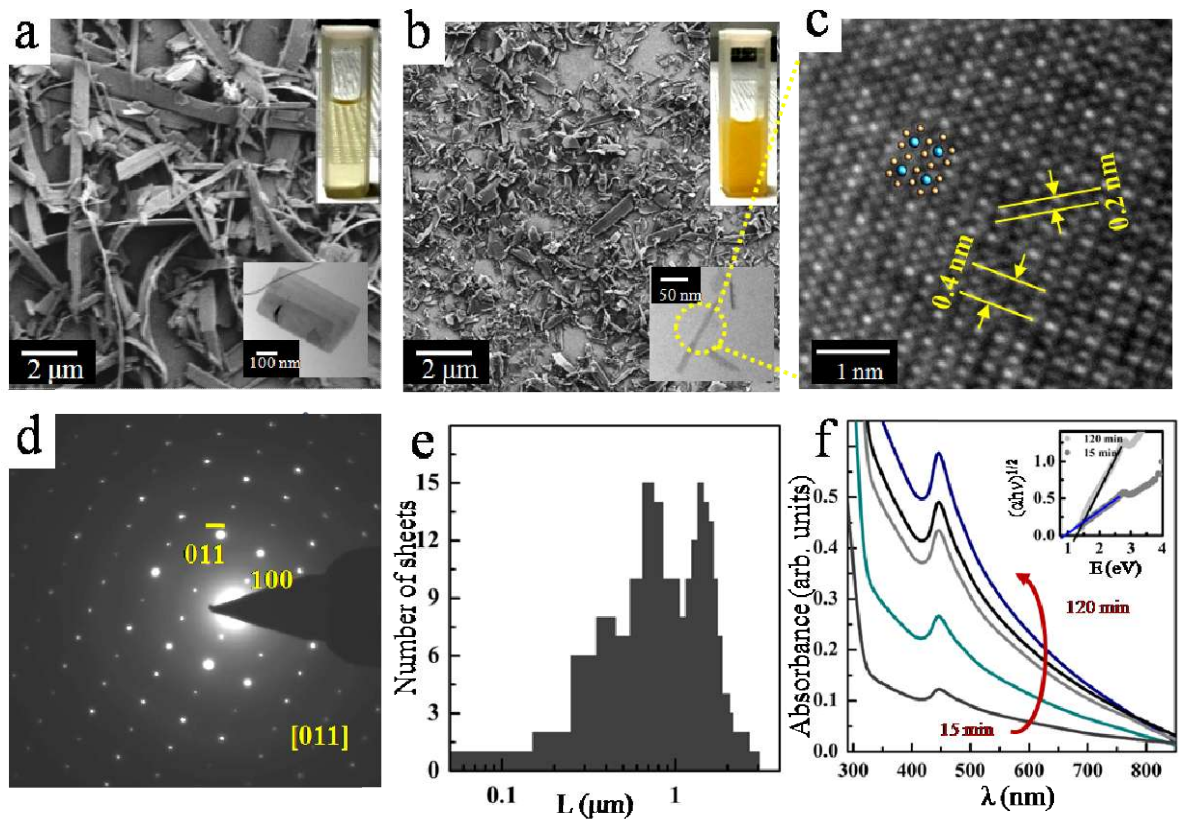

**Figure 3. Characterization of exfoliated HfS2 nanosheets.** (a) FESEM image of the nanosheets after 15 min of ultrasonication in CHP confirming the beginning of exfoliation process. Inset represents the photograph of suspension after 15 min of exfoliation and the respective TEM image of nanosheets revealing the presence of few layers. (b) FESEM image of the nanosheets after 120 min of ultrasonication in CHP resulting in the formation of ultra-thin layers of HfS2 nanosheets. Inset represents the photograph of suspension after 120 min of exfoliation and the respective TEM image of ultra-thin sheets confirming the completion of exfoliation process. (c) Atomic scale HRTEM image from an ultra-thin sheet showing the top view HfS2 hexagonal structure (*Hf* atoms: blue spheres, *S* atoms: orange spheres) (d) Selected-area electron diffraction pattern. (e) Statistical distribution of the size of ultra-thin nanosheets from FESEM image (sample size: 180 sheets). (f) UV- Visible absorption spectroscopy of the suspension of HfS2 in CHP as a function of sonication time. Inset shows the calculated indirect-bandgap for 15 min and 120 min sonicated suspension.

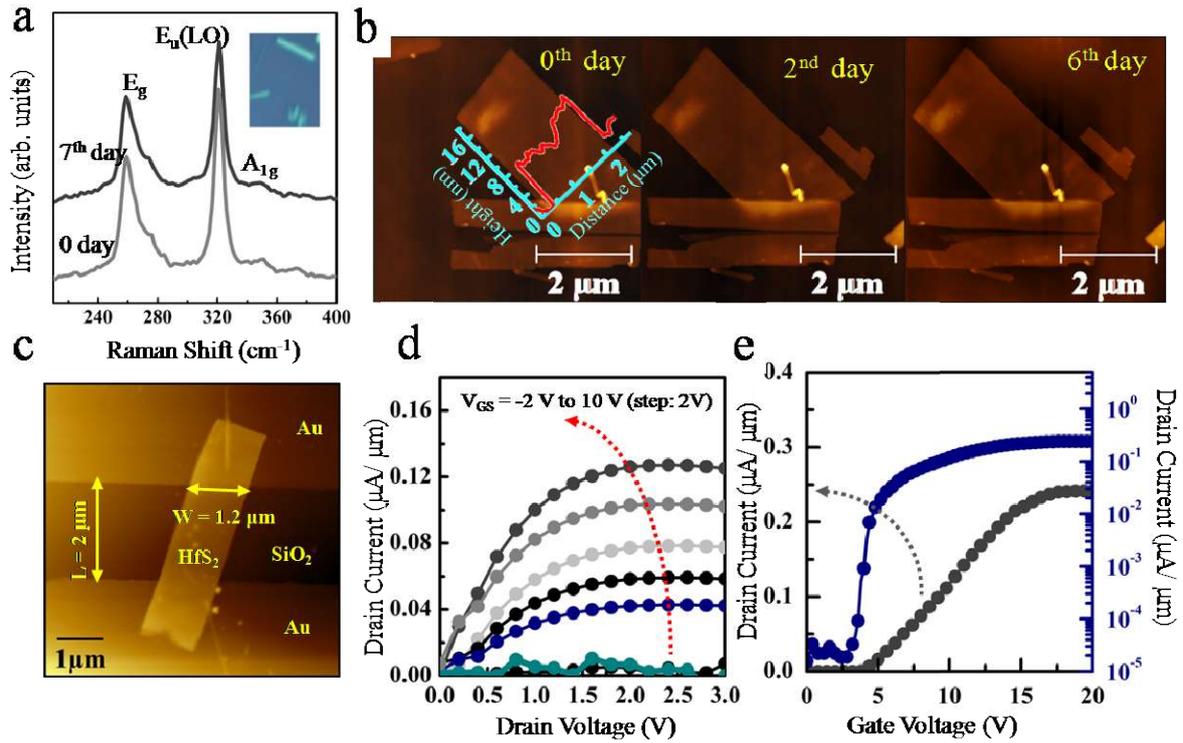

**Figure 4. Ambient stability of the nanosheets and device characteristics of HfS2-FET.** (a) Comparison of Raman spectra taken on the first day and after a weak of ambient exposure confirming the stability of nanosheets. Inset shows the optical image. (b) AFM images of a few layer HfS2 flake (Inset shows the height profile revealing a thickness of 8 nm) continuously scanned for a week confirming the absence of bubble formation and the sharpness of the AFM images signifying there persistence against ambient degradation. (c) AFM of the FET device (W: channel width, L: channel length, thickness of the HfS2 flake: 18 nm). (d) Carrier transport by HfS2 as a function of gate voltage (VGS). (e) Transfer characteristics of the HfS2-FET (left y-axis: linear scale, right y-axis: logarithmic scale).

ELECTRONIC SUPPLEMENTARY MATERIAL

# High yield synthesis and liquid exfoliation of two-dimensional belt like hafnium disulphide

*Harneet Kaur[1,*], Sandeep Yadav[2,§], Avanish K. Srivastava[1], Nidhi Singh[1], Shyama Rath[3], Jörg J. Schneider[2,§], Om P. Sinha[4] and Ritu Srivastava[1,*]*

[1]National Physical Laboratory, Council of Scientific and Industrial Research, Dr. K. S. Krishnan Road, New Delhi 110012, India.

[2]Technische Universität Darmstadt, Eduard-Zintl-Institut für Anorganische und Physikalische Chemie L2 I 05 117, Alarich-Weiss-Str 12, 64287 Darmstadt, Germany.

[3]Department of Physics and Astrophysics, University of Delhi, Delhi 110007, India.

[4]Amity Institute of Nanotechnology, Amity University UP, Sector 125, Noida, Uttar Pradesh 201313, India.

*address for correspondence: harneet@mail.nplindia.org
ritu@nplindia.org
§Material synthesis and characterization

**1 Atomic force microscopic image of micromechanically cleaved HfS$_2$ flake**

The as-prepared crystals of HfS2 were micromechanically cleaved using scotch-tape and pasted onto cleaned silicon substrate. The adhesive of the tape on substrate were then removed using ultrasonication in acetone and propanol followed by baking the substrate at 120$^o$C for 2 hours in vacuum (10-5 mbar) prior to AFM measurements. The AFM image of the sample was taken in tapping mode. The thickness of the flake was found to be 300 nm.

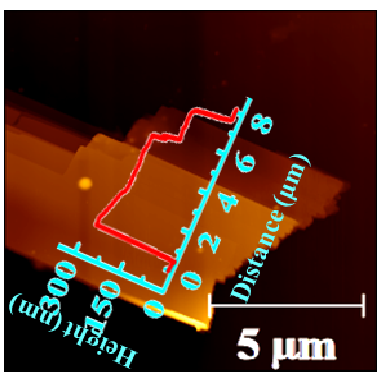

**Figure S1** AFM image of HfS2 flake on silicon wafer. The graph in the inset represents the height profile.

## 2 Thickness of exfoliated HfS₂ as a function of centrifuge speed

The thickness of the exfoliated HfS2 nanosheets in CHP was found out using atomic force microscopy on drop-cast samples prepared by different supernatant centrifuged at different angular speeds. The as-sonicated suspension of HfS2 (120 minutes, CHP) is subjected to centrifugation to separate the nanosheets of different thickness as shown in the schematic in Figure 2a. The suspension centrifuges at 1000 r.p.m. (60 min, P1) and supernatant is decanted while the sediment consisting of un-exfoliated HfS2 is discarded. Then the supernatant P1 is subjected to a second high speed centrifugation at 13,000 r.p.m. (30 min, P2) to separate out a supernatant of ultra-thin HfS2 nanosheets. The sediment remaining after this process is mixed with fresh CHP by mild agitation and subjected at a lower r.p.m of 10,000 r.p.m (30min, P3) and 4,000 r.p.m (30 min, P4), repeating same multiple steps to achieve nanosheets of different thicknesses. Figure 2b represents the AFM of the nanosheets centrifuged at 1000 r.p.m (60 min, P1) shows a large distribution of height from 0 to 80 nm as seen in the AFM apparent height scale. The AFM of supernatant P2 (Figure 2c) shows the presence of ultra-thin nanosheets of HfS2, minimize the window of apparent height in AFM image from 0 to 5 nm. Figure 2d represents the AFM of the supernatant P3 (10,000 r.p.m, 30 min) shows relatively large lateral dimensions of nanosheets compared to P2 having almost same apparent height distribution from 0 to 6 nm. The AFM of supernatant P4 (4000 r.p.m, 30 min) consists of flat sheets having lateral dimensions of the order of 1-4 microns and height profile of the flakes ranges from 10-30 nm (Figure 2d).

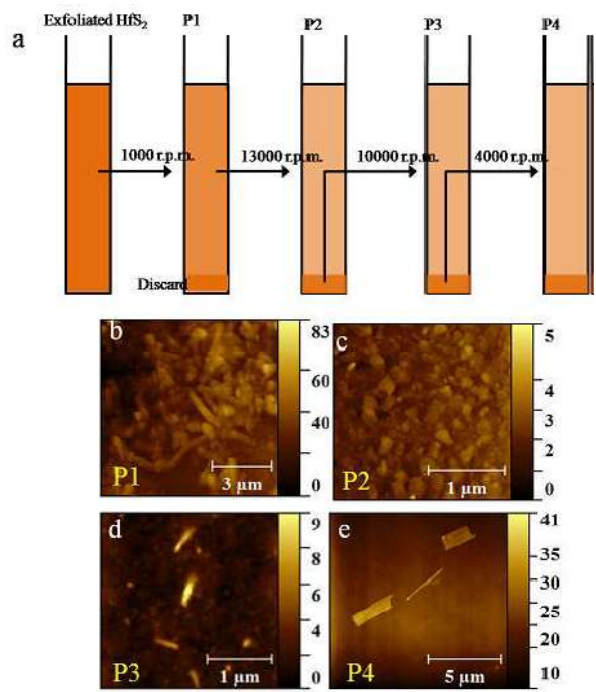

**Figure S2** (a) Schematic representation of the protocol of centrifugation to separate out the nanosheets of different thickness from the as-sonicated dispersion. (b,c,d,e) AFM images of the different samples on silicon wafer prepared by drop-cast of supernatants centrifuged at different r.p.m.

## 3 Thickness distributions of exfoliated HfS$_2$ nanosheets

The thickness of the exfoliated nanosheets was measured using AFM. Figure 3 represents the number of nanosheets versus thickness of the sample prepared by using P2 supernatant as discussed in section S2. The dominant thickness corresponds to 3 nm which is approximately 5-6 layers of HfS2.

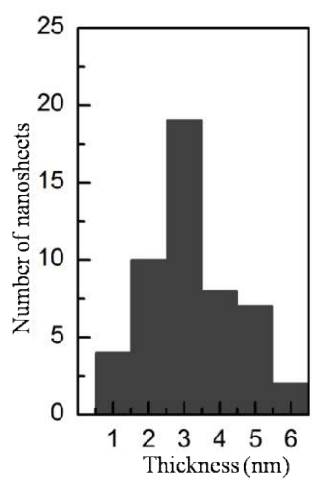

**Figure S3** Distribution of number of nanosheets versus thickness measured by the apparent height profile of the AFM images of the samples prepared by using supernatant centrifuged at 13,000 r.p.m (sample size = 50 sheets).

## 4 Degradation studies of HfS$_2$ in various solvents

Previous reports on HfS2 [1-2] has shown that the exfoliated nanosheets degrades during ambient exposure, limiting this material electronic applications. However, we found that liquid phase exfoliation is a reliable method as suitable solvents provides protection shell to the nanosheets, slowing down the degradation time. In this respect, HfS2 nanobelts were exfoliated in various solvents such as N-methyl-2-pyrrolidone (NMP), dimethylformamide (DMF) and CHP and the time dependent optical absorption over a period of 250 hours is studied. Figure 4a shows the comparative absorbance of the nanosheets of HfS2 sonicated in various solvents. However, over time, a decrease in the absorbance spectra as shown in

Figure 4b, c, d is observed, revealing the degradation of nanosheets over time resulting in loss of optical absorbance in all the three solvents. But CHP is found to offer low reactivity compared to NMP and DMF as observed in Figure 4e, f, and g representing the absorbance at a wavelength of 445 nm as a function of time. Fitting these curves with exponential decays using equation: $A = A_1 + A_2 e^{-t/\tau}$ similar to other research groups [3-4], we found decay constant ($\tau$) to be the largest in case of CHP ($\tau = 605 \pm 21$ h) compared to NMP ($\tau = 305 \pm 4$ h) and DMF ($\tau = 244 \pm 3$ h), confirming the nanosheets to be most stable in CHP solvent. However, more work is needed in this in this direction to expand the list of solvents in this direction.

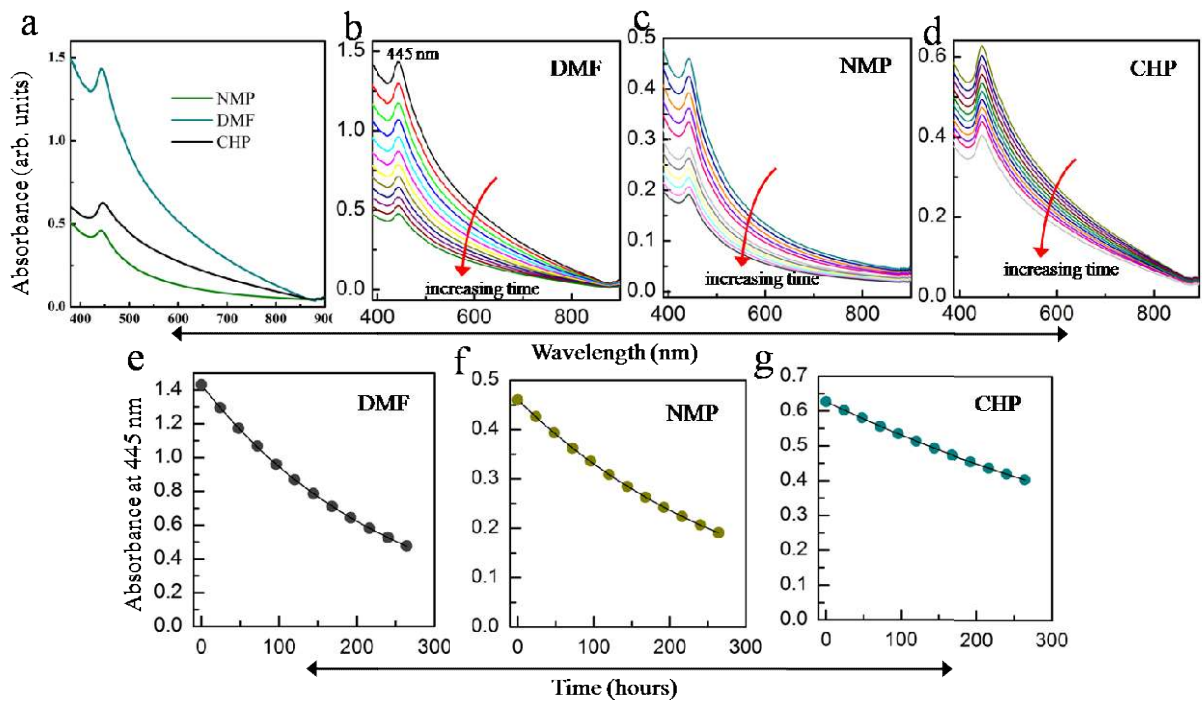

**Figure S4 Degradation kinetics by UV-Vis spectroscopy.** (a) absorbance of HfS2 suspension in various solvents (b,c,d) absorbance spectra of HfS2 dispersion in DMF, NMP and CHP respectively over time showing nanosheets degradation (e,f,g) absorbance at 445 nm, measured as a function of time for HfS2 dispersion exfoliated in DMF, NMP and CHP respectively. The dots represent the experimental data and the black line represents exponential decays.

## 5 FET characteristics of a thick layered HfS$_2$

The thickness dependent electronic properties of HfS2 has been explored by studying the transistor characteristics of a 110 nm layered flake of HfS2 as a conducting channel material between source and drain. The AFM image of the device is shown in Figure 5a. A clear saturation of drain current (IDS) was observed in Figure 5b corresponding to each gate bias voltage which was tuned from -5V to 25 V. However, in case of thin flake, saturation of drain current was achieved at sufficiently low gate voltage (Figure 4 in manuscript). Furthermore, the transfer characteristic in Figure 5c for a fixed VDS of 5V reveals uni-polar n-type transport behavior similar to thin few layered flake. Thus, regardless of thickness of the HfS2, the electrical conductivity shows n-type behavior. A maximum drain current of 0.04 µA/µm is obtained at a gate voltage of 10 V. The FET device exhibits a poor current on-off modulation ratio of 40 at room temperature when gate voltage is varied from 0V to 10 V which is extremely low compared to the thin $HfS_2$ flake, revealing thicker devices of $HfS_2$ shows similar electrical behavior like other 2D materials.

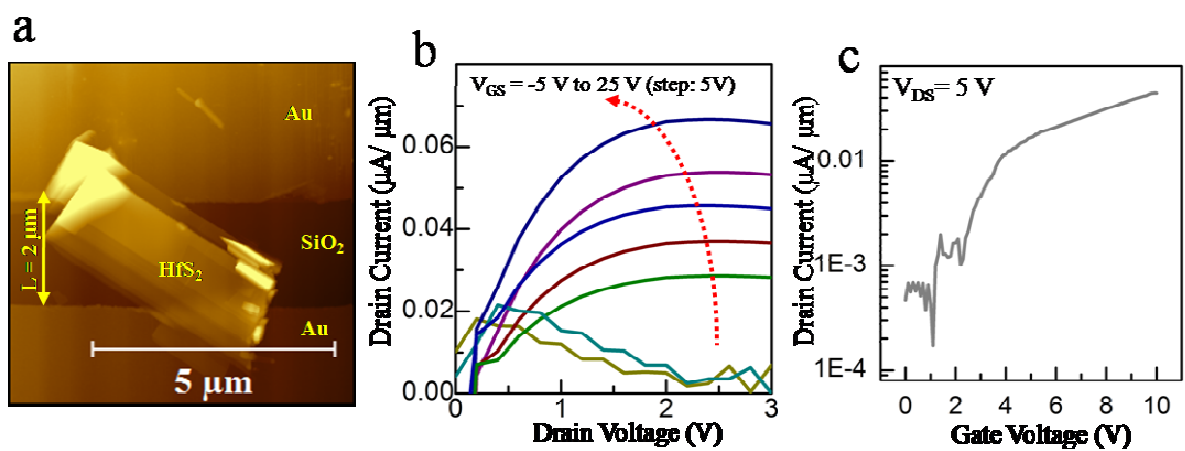

**Figure S5** (a) AFM image of a 110 nm thick layered HfS2 flake connecting source and drain. (b) Carrier transport characteristic of FET as a function of gate voltage (VGS). (e) Transfer characteristics of the HfS2-FET (y-axis: logarithmic scale).